\documentclass[aps,pre,preprint,superscriptaddress,showpacs,showkeys]{revtex4-1}
\usepackage{amssymb}
\usepackage{amsmath}
\usepackage{amsfonts}
\usepackage{graphicx}
\usepackage{booktabs} % For formal tables
\usepackage{newfloat,algcompatible}
\usepackage{tabularx}
\usepackage{pdfpages}
\DeclareFloatingEnvironment[
    fileext=loa,
    listname=List of Algorithms,
    name=ALGORITHM,
    placement=tbhp
]{algorithm}
\usepackage{enumerate}

\makeatletter
\AtBeginDocument{\let\LS@rot\@undefined}
\makeatother

\begin{document}
\title{Link transmission centrality in large-scale social networks}

\author{Qian Zhang}
\affiliation{Laboratory for the Modeling of Biological and Socio-technical Systems, Northeastern University, Boston MA 02115 USA}
\author{M\'arton Karsai}
\affiliation{Laboratoire de l'Informatique du Parall\'elisme, INRIA-UMR 5668, IXXI,  ENS de Lyon, 69364 Lyon, France}
\affiliation{Laboratory for the Modeling of Biological and Socio-technical Systems, Northeastern University, Boston MA 02115 USA}
\author{Alessandro Vespignani}
\affiliation{Laboratory for the Modeling of Biological and Socio-technical Systems, Northeastern University, Boston MA 02115 USA}
\affiliation{Institute for Scientific Interchange Foundation, Turin 10133, Italy}
%\affiliation{Institute for Quantitative Social Sciences at Harvard University, Cambridge, MA, 02138 USA} 

\date{\today} \widetext

\begin{abstract} 
Understanding the importance of links in transmitting information in a network can provide ways to hinder or postpone ongoing dynamical phenomena like the spreading of epidemic or the diffusion of information. In this work, we propose a new measure based on stochastic diffusion processes, the \textit{transmission centrality}, that captures the importance of links by estimating the average number of nodes to whom they transfer information during a global spreading diffusion process. We propose a simple algorithmic solution to compute transmission centrality and to approximate it in very large networks at low computational cost. Finally we apply transmission centrality in the identification of weak ties in three large empirical social networks, showing that this metric outperforms other centrality measures in identifying links that drive spreading processes in a social network.
\end{abstract}

%\pacs{0000,0000}

\maketitle

\section{Introduction}
\label{intro}
The importance of nodes and links in networks is commonly measured through centrality measures. Their definitions generally rely on local and/or global structural information. Centrality measures using local information, like the node degree or link overlap, are computed efficiently as they only require knowledge about the neighbors of a given node or link. On the other hand, these measures cannot provide information on which nodes or links play global roles in the network structure. On the contrary, centrality measures based on global information about the network structure, like betweenness and closeness  centrality~\cite{Freeman1977A,Bavelas1950Communication}, Katz centrality~\cite{Katz1953A}, k-shell index~\cite{Bollobas1984Graph,Kitsak2010Identification}, subgraph centrality~\cite{Estrada2005Subgraph} and induced centrality measures~\cite{Everett2010Induced} may better characterize the overall importance of a node or link. Unfortunately, although effective algorithms for approximating these quantities have recently been proposed~\cite{brandes2001faster,Ercsey2010Centrality}, estimating these measures in large scale networks is still computationally challenging.

While global centrality measures have been very successful in identifying structurally important nodes or link in networks, it has been argued~\cite{Borgatti2004Centrality} that they do not evidently identify nodes or links with a key role in dynamical processes. Other centrality metrics, which directly use dynamical processes to assign importance were found to be more successful in this sense. The best examples are metrics based on random walkers like PageRank \cite{Brin1998The}, eigenvector centrality \cite{Leontief1941Structure}, or accessibility \cite{Travencolo2008Accessibility}. Other examples are local metrics like the expected force \cite{Lawyer2015Understanding}, or percolation centrality \cite{Piraveenan2013Percolation}. These measures are based on random diffusion processes, but do not fully capture the basic mechanisms behind contagion mediated spreading phenomena. Here we define a new link centrality measure, \textit{transmission centrality}, tailored to identify the role of nodes and links in controlling contagion phenomena. The transmission centrality measures the average number of nodes who are reached by the contagion process through each link during the spreading of a stochastic contagion process. This provides a direct measure of the centrality of the link in hindering or facilitating the contagion process. In the case of very large-scale network, we propose a heuristic calculation of transmission centrality, which is both computationally efficient and can be easily extended for weighted, directed, or temporal networks or even for nodes. Furthermore, to demonstrate the usefulness of transmission centrality we present a case study where we use this metric to identify weak ties~\cite{Granovetter73,Granovetter83} in social networks and characterize their role in contagion processes.

As it follows, after a brief discussion of related works and utilized datasets, we formally introduce transmission centrality and discuss a heuristic method for its approximate calculation. Then we discuss its properties and correlations with local centrality measures in three large-scale real world social networks. Finally, we present simulation results of SIR spreading processes to demonstrate the capacity of combined local measures and transmission centrality in designing effective strategies to enhance or hinder information diffusion in social networks.

\section{Related works}
Node centralities have been widely studied, from classical static centralities like degree, closeness, betweenness, eigenvector \cite{newman2010networks} to centrality measures based on dynamical processes, such as random walk (e.g. PageRank \cite{Brin1998The}). Among these, betweenness centrality is one of the most popular measures as it quantifies the importance of a node by considering the global structure of a network instead of local information. Unfortunately, the efficiency of algorithms to calculate betweenness centrality is still challenging in the case of large-scale social networks as its best computation method has $\mathcal{O}(|V||E|)$ complexity for unweighted networks and $\mathcal{O}(|V||E|+|V|^{2}\log |V|)$ for weighted networks \cite{brandes2001faster}. While many variants and approximation algorithms have been proposed to improve its algorithmic efficiency \cite{brandes2007centrality,brandes2008variants,bader2007approximating,geisberger2008better,riondato2016fast,Jensen2016Detecting}, researchers have also proposed alternative measures to quantify the importance of nodes in terms of dynamical processes on top of a network, such as K-path centrality \cite{alahakoon2011k} and percolation centrality \cite{Piraveenan2013Percolation}. K-path centrality \cite{alahakoon2011k} applies self-avoiding random walks of length $k$ and counts the probability that a message originating from a given source traverses a node as its centrality. The percolation centrality \cite{Piraveenan2013Percolation} measures the relative importance of a node based on both network structure and its percolated states. Single-node-influence centrality and Shapley centrality assess the importance of a node in isolation and in a group respectively in social influence propagation processes \cite{chen2017interplay}. \cite{rossi2016exploring} simulates epidemic models (SIS and SIR) to estimate node centralities on top of temporal social networks. Interestingly, this study shows that spreading processes fail to characterize the centrality measures like degree and core numbers of infected nodes. Dynamics-sensitive centrality \cite{liu2016locating}, which counts the outbreak size in an epidemic model to quantify spreading influence of nodes, can better capture the importance of nodes particularly in epidemic spreading processes.

Most centrality algorithms have also been generalized to the estimation of link centrality measures, such as edge betweenness centrality, spanning edge betweenness centrality \cite{teixeira2013spanning,mavroforakis2015spanning}, and K-path edge centrality \cite{de2012novel}. As node centralities aim to characterize the importance of nodes in a network, edge centralities provide quantitative perspectives to measure the importance of links in a network structure \cite{de2014facebook,everett2016bridging}.

\section{Materials and Methods}

\subsection{Network Data Descriptions}
\label{sec:data}

In the following study, we will discuss centrality algorithms by using three distinct sets of data recording communications between thousands or millions of individuals. For each dataset, first we aggregate the sequence of interactions to a static social network, excluding possible commercial communications. To do so, we only draw links between individuals who had at least one pair of mutual interactions during the observation period. In addition, to avoid leaf links we extract the $k$-core ($k=2$) structure \cite{Seidman1983,Bollobas1984} of each network and use their largest connected component (LCC). 

The first dataset we investigate is collected from the mobile phone call (MPC) communication sequences of $4,256,137$ individuals during 4 weeks with 1 second resolution \cite{Karsai2011,Kivela2012}. Individuals are anonymous users of a single operator with $20\%$ market share in a European country. The static social network contains $5,279,169$ mutual links. The final $k$-core ($k=2$) structure of the LCC includes $1,926,787$ nodes and $3,269,634$ edges.

The second social network is aggregated from the sequence of wall posts of Facebook users (FB) \cite{konect:2014:facebook-wosn-wall,viswanath09,konect}. The data records interactions from September 2004 to January 2009 between $31,720$ users connected by $80,592$ mutual links. The $k$-core ($k=2$) structure of the LCC of this network contains $20,244$ nodes and $70,132$ edges.

The last social network, A Twitter conversation network (TW), is constructed from tweets from October 2010 to November 2013, which were collected through the Twitter Gardenhose \cite{TwitterAPI}. We restrict our dataset to tweets with live GPS coordinates providing us over $420$ million communication events, which represent a $1-2\%$ of the entire volume. We construct a social network based on mutual conversational tweets (\textit{@mentions}) between $4,155,700$ users connected by $6,506,519$ links. The $k$-core ($k=2$) structure of the LCC of the Twitter conversation network consists of $966,779$ nodes linked by $2,779,524$ edges.

\subsection{Transmission centrality}
\label{sec:definition}

\begin{algorithm}[h!]
\begin{algorithmic}[1]
%\scriptsize 
\REQUIRE $G=(V,E)$, $\beta$, $s$
\ENSURE $G_{BT}=(V_{BT},E_{BT})$ \textit{branching tree of spreading}
\vspace{.1in}
\State{$Q=queue()$	\hspace{.05in} // queue of $I$ nodes with susceptible neighbors}
\State{$G_{BT}.V_{BT}=\emptyset$ 	\hspace{.15in} // the branching tree}
\State{$G_{BT}.E_{BT}=\emptyset$}
\FOR{each vertex $u \in G.V-\{s\}$}
\State{$u.state$=$S$}
\State{$u.asc$=$NIL$}
\ENDFOR
\State{$s.state$=$I$}
\State{ENQUEU($Q$,$s$)}
\WHILE{$Q\neq \emptyset$}
\State{u=DEQUEUE(Q)}
\State{SN=False	\hspace{.15in} // remaining susceptible neighbor of node $u$}
\FOR{each $v\in G.Adj[u]$}
\IF{$(v.state==S)$}
\IF{$(rand() \leq \beta)$}
\State{$v.state=I$}
\State{$u.asc=u$}
\State{$G_{BT}.V_{BT}.add(v)$}
\State{$G_{BT}.E_{BT}.add((u,v))$}
\State{ENQUEUE(Q,v)}
\ELSE
\State{SN=True}
\ENDIF
\ENDIF
\IF{SN==True}
\State{ENQUEUE(Q,u)}
\ENDIF
\ENDFOR
\ENDWHILE
\end{algorithmic}
\caption{Susceptible-Infected process}
\label{alg:SI}
\end{algorithm}

\begin{figure*}
	\begin{center}
		\includegraphics[width=1.0\textwidth]{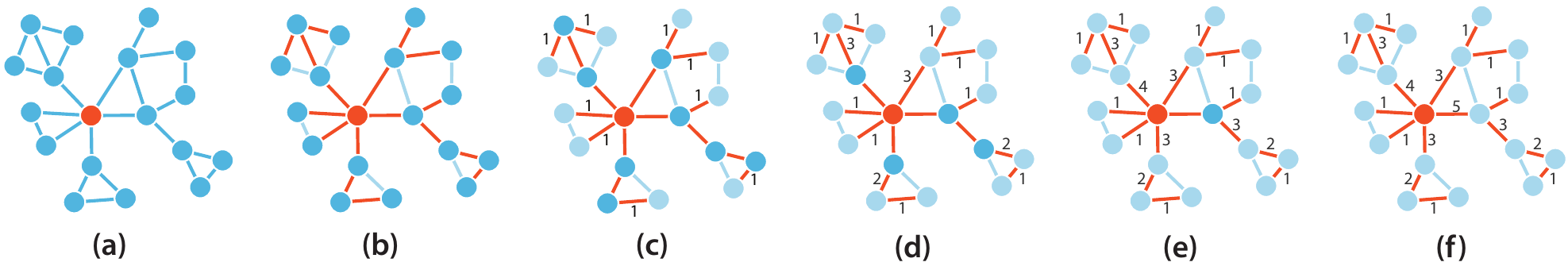}
	\end{center}
\caption{Calculation of \textit{transmission centrality} of links. (a) A network with a randomly selected seed node; (b) the branching tree rooted from the initial seed (root and edges in the tree are colored in red); (c) for each leaf edge in the branching tree increase the counter by 1; (d)-(f) remove leafs and increase the counter of their ascendant by the counter of the removed leafs.}
\label{fig:riverbasin}
\end{figure*}

\textit{Transmission centrality} aims to measure for each link in a network its influence in disseminating some globally spreading information. More precisely it measures the number of nodes who received information during a diffusion process through a given link. Its definition intrinsically assumes a diffusion process to unfold on a network structure. In our definition we use the simplest possible information spreading process, the \textit{Susceptible-Infected model}~\cite{Barrat2008}, however this can be replaced by any other diffusion process. The Susceptible-Infected (SI) process is defined on a connected network $G=(V,E)$, where nodes $u\in G.V$ can be in two mutually exclusive states, either susceptible ($S$) or infected ($I$). Initially each node is susceptible ($S$) except a randomly selected seed node, which is set to be in state $I$. In one iteration step each infected node can infect its susceptible neighbors with rate $\beta$ until every node becomes infected in the network. Note that the parameter $\beta$ here scales with the speed of information spreading, with value $\beta=1$ corresponding to the fastest possible information diffusion process determining the shortest diffusion routes between the seed and any other node in the network. This diffusion process can be simulated with a modified breath-first-search algorithm~\cite{Cormen2001Introduction} as written in Alg.\ref{alg:SI}. There, during the unfolding of the diffusion we keep infected nodes with susceptible neighbors in a $Q$ queue and record the branching tree $G_{BT}=(V_{BT},E_{BT})$ of the process by keeping track of the direct ascendant of each node from which it received the information. Exploiting the structure of the actual branching tree, \textit{transmission centrality} is formally defined as
\begin{equation}
C_{tr}(u,v) = 
\begin{cases}
   \max(|desc(u)|,|desc(v)|), & \text{if } (u,v)\in E_{BT}, \\
   0, & \text{otherwise}
\end{cases}
\end{equation}
where $|desc(i)|$ denotes the number of descendant nodes of node $i$ in the branching tree of the actual spreading.

The branching tree $G_{BT}$, which is a subgraph of $G$, encodes the diffusion paths that the information takes to reach the vertices of the network. Using its structure we can easily measure the actual $C_{tr}$ value of each link by performing a second step of calculation based on the \textit{river-basin algorithm}~\cite{Iturbe2001}. In practice, taking the initial seed $s$ as the root of $G_{BT}$, and starting from the leafs of the branching tree we can count the number of descendant nodes of each link, i.e., who received the information via the actual link. The algorithm is summarized in Alg.\ref{alg:riverbasin}, illustrated in Fig.\ref{fig:riverbasin} and works as follows:

\begin{algorithm}[h!]
\begin{algorithmic}[1]
%\scriptsize
\REQUIRE $G=(V,E)$ and $G_{BT}=(V_{BT},E_{BT})$
\ENSURE $C_{tr}$ dictionary of \textit{transmission centrality} values
\State{$C_{tr}=dict()$}
\FOR{$(u,v) \in G.E$}
%\State{initialize a counter $c_{u,v}=0$}
\State{$C_{tr}((u,v))=0$   \hspace{.4in} // set counter to zero for each link}
\ENDFOR
\WHILE{$G_{BT}.E_{BT} \neq \emptyset $}{
\FOR{$ v \in G_{BT}.V_{BT}$}
\IF{$deg(v)==1$}
		\State{$p=asc(v)$	\hspace{.4in} // parent node of v}
		\State{$gp=asc(p)$	\hspace{.4in} // grandparent node of v}
	  	\State{$C_{tr}((v,p)) = C_{tr}((v,p))+1$}
	  	\State{$C_{tr}((p,gp)) = C_{tr}((p,gp))+C_{tr}((v,p))$}
	    \State{$G_{BT}.E_{BT} \longleftarrow G_{BT}.E_{BT} - \{ (v,p) \}$}
	    \State{$G_{BT}.V_{BT} \longleftarrow G_{BT}.V_{BT} - \{ v \}$}
\ENDIF
\ENDFOR
	}
\ENDWHILE
\end{algorithmic}
\caption{Transmission centrality}
\label{alg:riverbasin}
\end{algorithm}

First we define a dictionary $C_{tr}$, which associates a counter to each link $(i,j)\in G.E$, that we set to zero initially (lines 1-3 in Alg.\ref{alg:riverbasin}). Then we recursively do the following for every node $v\in G_{BT}.V_{BT}$, which appears with degree $deg(v)=1$ in $G_{BT}$:
\begin{enumerate}[(a)]
\item Increase by one the counter $C_{tr}((v,p))$ of the (leaf) edge $e_f=(v,p)\in G_{BT}.E_{BT}$, which connects $v$ to its parent node $p=asc_{BT}(v)$ in $G_{BT}.V_{BT}$ (line 10 in Alg.\ref{alg:riverbasin}).
\item Increase by $C_{tr}((v,p))$ the counter $C_{tr}((p,gp))$ of its ascendant edge $asc_{BT}(e_f)=(p,gp)$, where $gp=asc(p)$ is the grandparent node of $v$ in $G_{BT}.V_{BT}$ (line 11 in Alg.\ref{alg:riverbasin}).
\item Remove $v$ from $G_{BT}.V_{BT}$ and $e_f$ from $G_{BT}.E_{BT}$ (line 12 and 13 in Alg.\ref{alg:riverbasin}). The final transmission transmission centrality value of the actual link $e_f=(v,p)$ is stored in $C((v,p))$.
\end{enumerate}
By repeating II.(a)-(c) recursively for each appearing leaf edge we assign a non-zero value for each link in the branching tree as it is demonstrated in Fig.\ref{fig:riverbasin}.c-f.

\begin{algorithm}[h!]
\begin{algorithmic}[1]
%\scriptsize
\REQUIRE $G=(V,E), \beta$
\ENSURE $C^{avr}_{tr}$ dictionary of \textit{average transmission centrality} values
\State{$C^{avr}_{tr}=dict()$}
\FOR{$(u,v) \in G.E$}
\State{$C^{avr}_{tr}((u,v))=0$   \hspace{.4in} // set counter to zero for each link}
\ENDFOR
\FOR{$v \in G.V$}
	\State{$G_{BT} \longleftarrow SusceptibleInfected(G,\beta)$}
	\State{$C^{act}_{tr} \longleftarrow TransmissionCentrality(G, G_{BT})$}
	\FOR{$(u,v) \in G.E$}
		\State{$C^{avr}_{tr}((u,v))+=C^{act}_{tr}((u,v))$ \hspace{.2in} // summing realisations}
	\ENDFOR
\ENDFOR
\FOR{$(u,v) \in G.E$}
\State{$C^{avr}_{tr}((u,v))=C^{avr}_{tr}((u,v))/|G.V|$ \hspace{.12in} // computing averages}
\ENDFOR
\end{algorithmic}
\caption{Average transmission centrality}
\label{alg:avr}
\end{algorithm}

The transmission centrality of a link can take values between 0 (for links, which are not in the branching tree) and $(N-1)$ (e.g. in the case the seed is propagating information via a single link). Its actual value depends on the choice of the seed node and on the structure of the branching tree determined by the stochastic diffusion process. In this way centrality values of the same link may vary from one realization to another. To eliminate the effects of such fluctuations the final definition of transmission centrality of links is taken as the average centrality value for each link computed over processes initiated from every node in the network (for a algorithmic definition see Alg.\ref{alg:avr}). Note that from now on $C_{tr}$ always assigns an average quantity if not stated otherwise.

\section{Results}
\subsection{Heuristic calculation of transmission centrality}

One iteration to measure $C_{tr}$ performs with $\mathcal{O}(|E|)$ time complexity, in this case where we initiate its calculation from every node $v\in V$, its overall complexity is $\mathcal{O}(|V||E|)$. 
It is however possible to define a heuristic estimate of transmission centrality at a considerably small computational cost. As the branching trees of different realizations may largely overlap, a relatively small number of independent realizations, initiated from a reduced set of randomly selected seeds, could provide a good approximation to transmission centrality. Link transmission centrality initiated from a single node provides a locally biased measure as it assigns higher values to links closer to the actual seed. This bias is averaged out if we initiate the spreading process from every node in the network, but in case of a limited number of seeds it has  residual effects. One way to eliminate this residual bias is by assigning zero centrality values to links connecting nodes closer than a distance $d$ to the actual seed. The best value of $d$ depends on the network; however this can be estimated by parameter scanning, as demonstrated in SI Fig. S1. 

\begin{figure}[ht!]
\begin{center}
		\includegraphics[width=0.65\textwidth]{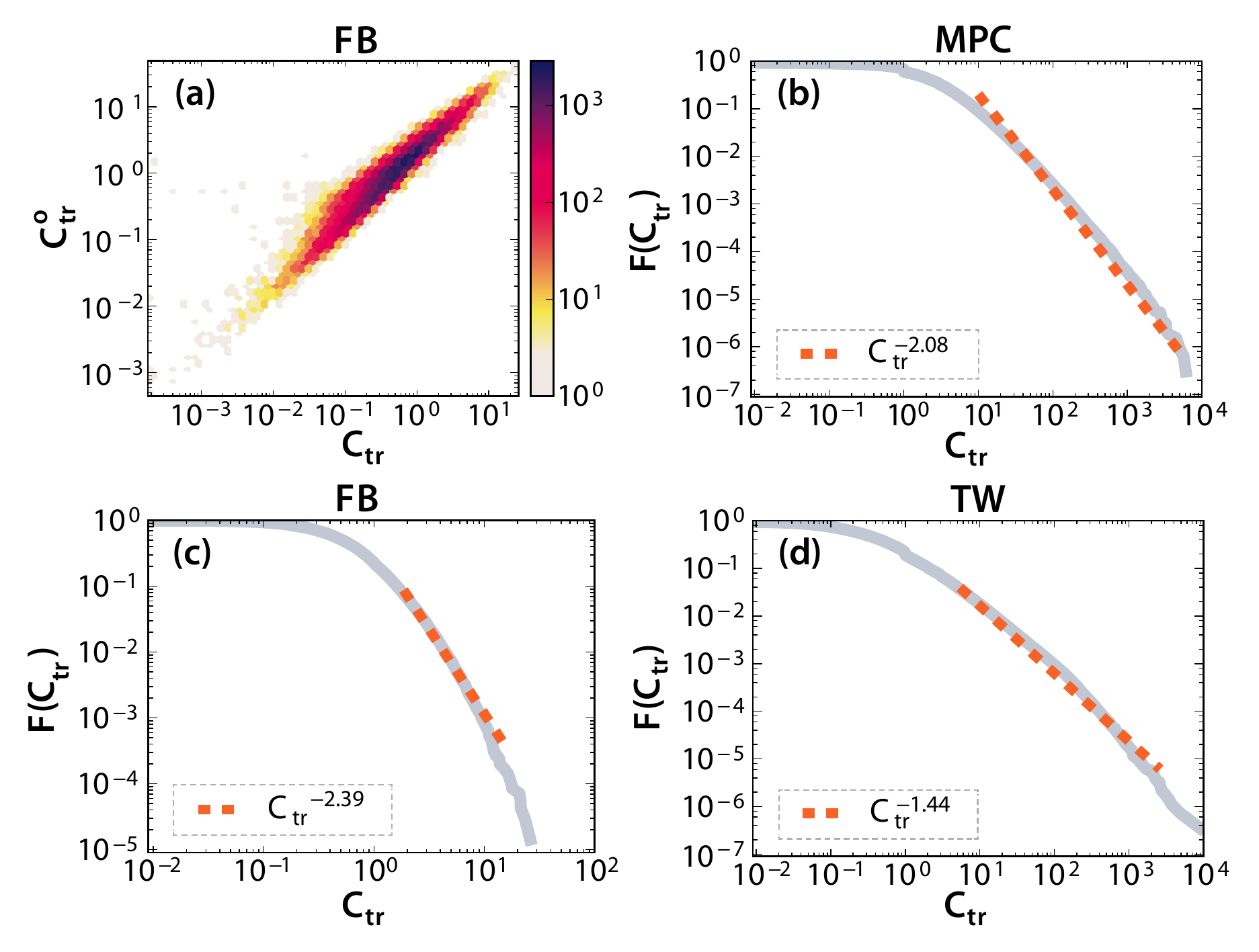}
 \end{center}
 \caption{(a) Correlation heat-map plot between the $C_{tr}^{o}$ exact and approximated $C_{tr}$ transmission centrality values of the FB network. Approximate measures were initiated from $5000$ seeds and unbiased in $d=3$ distance. (b,c,d) $F(C_{tr}) = P(C_{tr}^{'} \leq C_{tr})$ cumulative distribution functions of transmission centrality values in the MPC, FB, and TW networks respectively. $C_{tr}$ values were averaged for each link from respectively $2000$, $5000$ and $5000$ random realizations with un-biasing distances $d=8$, $3$, and $7$. Power-law functions (dashed orange lines) fitted with exponents $\alpha \simeq 3.08$, $3.39$, and $2.44$ respectively.}
  \label{fig:ctrprop}
\end{figure}

To illustrate the computation of the heuristic estimate, we use the FB network with $20,244$ nodes (for more details see Section \ref{sec:data}) and compute the average transmission centrality for each link via the exact method by initiating an SI process from each node and the heuristic method where we initiate processes from $5000$ random seeds (i.e. $\sim 25\%$ of all nodes) and eliminate biases in distance $d=3$ around each seed (for more on the selection of this value see SI Fig. S1). In Fig.\ref{fig:ctrprop}a we present a  heat-map plot about the correlation between the exact (assigned as $C^{o}_{tr}$ here) and the approximated (assigned as $C_{tr}$) centrality values of each link. It is evident that there is a strong correlation between the exact and approximated values of centralities, quantified by an $r=0.96$ ($p<10^{-6}$) Pearson correlation coefficient. Consequently, this unbiased sampling method can provide very close approximations to the exact transmission centrality values, while considerably reducing the computational cost ($\sim 25\%$ in this case). Note that this correlation analysis was not repeated for the other two empirical networks as the computation of the exact method would take extremely long on such large networks due to its computational complexity.

Subsequently, we applied the approximate method to compute transmission centrality in the MPC network (with $2000$ seeds and $d=8$) and TW network (with $5000$ seeds and $d=7$) as well. We consistently found that the average unbiased transmission centrality of links, measured in the three empirical systems, are broadly distributed (see in Fig.\ref{fig:ctrprop}b-d respectively for the MPC, FB and TW networks) with power-law tails with exponents $\alpha=3.08$, $3.39$ and $2.44$ for the MPC, FB and TW networks respectively. This demonstrates the high variance of importance of links in transmitting information, which can be duly the consequence of the community rich structure of the three investigated social networks.

Transmission centrality can be generalized in various ways. First, it can be easily defined as a \textit{node centrality metric} by counting for each node the number of their descendant nodes in the branching tree. Moreover it can be extended for \textit{directed and/or weighted networks} by restricting the SI process to respect the direction of links during spreading or by scaling the transmission rate with the normalized weight of links. In addition, for an SI process one can explore central links in the case when the process does not diffuse along the shortest paths. By taking $\beta<1$, longer spreading paths become plausible allowing the inference of links, which are central in any scenario. Transmission centrality can be easily defined for \textit{temporal networks}~\cite{Holme2012Temporal} as well. Contrary to static networks, in temporal structures information can transmit between nodes only at the time of their interactions. As a result, information can travel only along time-respecting paths in the network, which drastically restricts the final outcome of any global contagion processes \cite{Karsai2014Time} and has evident consequences on the measured centrality values. Links, which appeared unimportant in the static structure may be central in the temporal network as they could lay on several time-respecting paths due to their specific interaction dynamics.

Finally, note that although transmission centrality is not equivalent, it naturally relates to the concept of betweenness centrality (and other centrality measures based on the counts of shortest paths between nodes). This relation is better explained in SI Fig. S2.

\subsection{Case study: weak tie identification to control contagion processes in social networks}

To demonstrate the potential of transmission centrality here we present a case study, where we use our new metric to identify ties in social networks in order to efficiently control contagion processes. Ties in social networks are associated with various strengths \cite{Onnela2007a,Saramaki2014,Palchykov2012} and commonly categorized into two mutually exclusive groups: weak and strong ties. Following the terminology introduced by Granovetter \cite{Granovetter73,Granovetter83}, weak ties are maintained via sparse interactions, bridging between tightly connected communities to keep the network connected \cite{Onnela2007a}, and play an important role in disseminating information globally \cite{Onnela2007b,Centola2007,Centola2010,Gao2013,Kossinets2006,Kumpula2007,Karsai2011,Miritello2011}. On the other hand strong ties, sustained by frequent communications, are crucial in shaping the local connectivity of social networks, they are responsible for emerging clustered topology \cite{Rapoport53,Kossinets2006,Kumpula2007}, and keeping information locally \cite{Centola2007,Centola2010,Karsai2011,Miritello2011}. A precise measure of tie strength would allow the efficient differentiation among these types and to identify weak ties in social networks in order to control globally spreading contagion processes.

\subsubsection{Conventional measures of social tie strengths.}
Several measures of social tie strength have lately been proposed in the literature, such as the link overlap
\begin{equation}
O(i,j)=\frac{n_{ij}}{(k_i-1)+(k_j-1)-n_{ij}},
\end{equation}
capturing the fraction of common friends in the neighborhood of connected nodes $i$ and $j$ \cite{Granovetter73,Onnela2007a,Onnela2007b}. Here, $k_{i}$ and $k_{j}$ assign the degree of node $i$ and $j$ respectively, and $n_{ij}$ is the number of their common neighbors. Weak ties are associated with small overlap values, while the contrary is not always true. Leaf links, structural holes, or merely the fact that networks are sparse may induce links with small overlap, which leads to some ambiguity when identifying weak ties in this way.

\begin{table}[ht!]
\centering
\caption{Pearson correlations between transmission centrality, overlap and dyadic tie strength}
\label{tab:correlations}
\begin{tabular}{l|ccc}
\hline
    & \multicolumn{3}{c}{Pearson Correlation ($p$-value)} \\
 network   & $(O, w)$     & $(C_{tr}, O)$     & $(C_{tr}, w)$    \\
\hline
MPC &  $0.097$ ($10^{-6}$) & $-0.126$ ($10^{-6}$) & $-0.023$ ($10^{-6}$) \\
FB  &  $0.151$ ($10^{-6}$) & $-0.148$ ($10^{-6}$) & $-0.098$ ($10^{-4}$) \\
TW  & $0.102$ ($10^{-6}$) & $-0.021$ ($10^{-6}$) & $-0.002$ ($10^{-3}$) \\
\hline
\end{tabular}
\end{table}

Another way to assign the strength of social ties is via the intensity of dyadic communication \cite{Onnela2007b,Barrat2004,Barrat2008}. It can be measured as the frequency, total duration, or the absolute number of interactions between connected peers. In this study, assuming discrete communication events, we define dyadic tie strength as the number of interactions between individuals $i$ and $j$ as
\begin{equation}
w(i,j)=\sum_{t=0}^{T} \delta(t,i,j)
\end{equation}
where the sum runs over the observation period $T$. $\delta(t,i,j)=1$ if an event appears between $i$ and $j$ at time $t$ regardless of its direction, otherwise it is $0$ \cite{Onnela2007b}. Dyadic tie strength may capture mutual commitment or emotional closeness between people; however, as a local measure, it is subjective to individual characteristics like communication capacity or the egocentric network size. In this way, it is unable to indicate the role of a link in the global structure in the context of the emergence of any collective phenomena. In addition its broadly distributed values prohibit an evident categorization of social ties. 

\begin{figure}[b!]
		\includegraphics[width=0.75\textwidth]{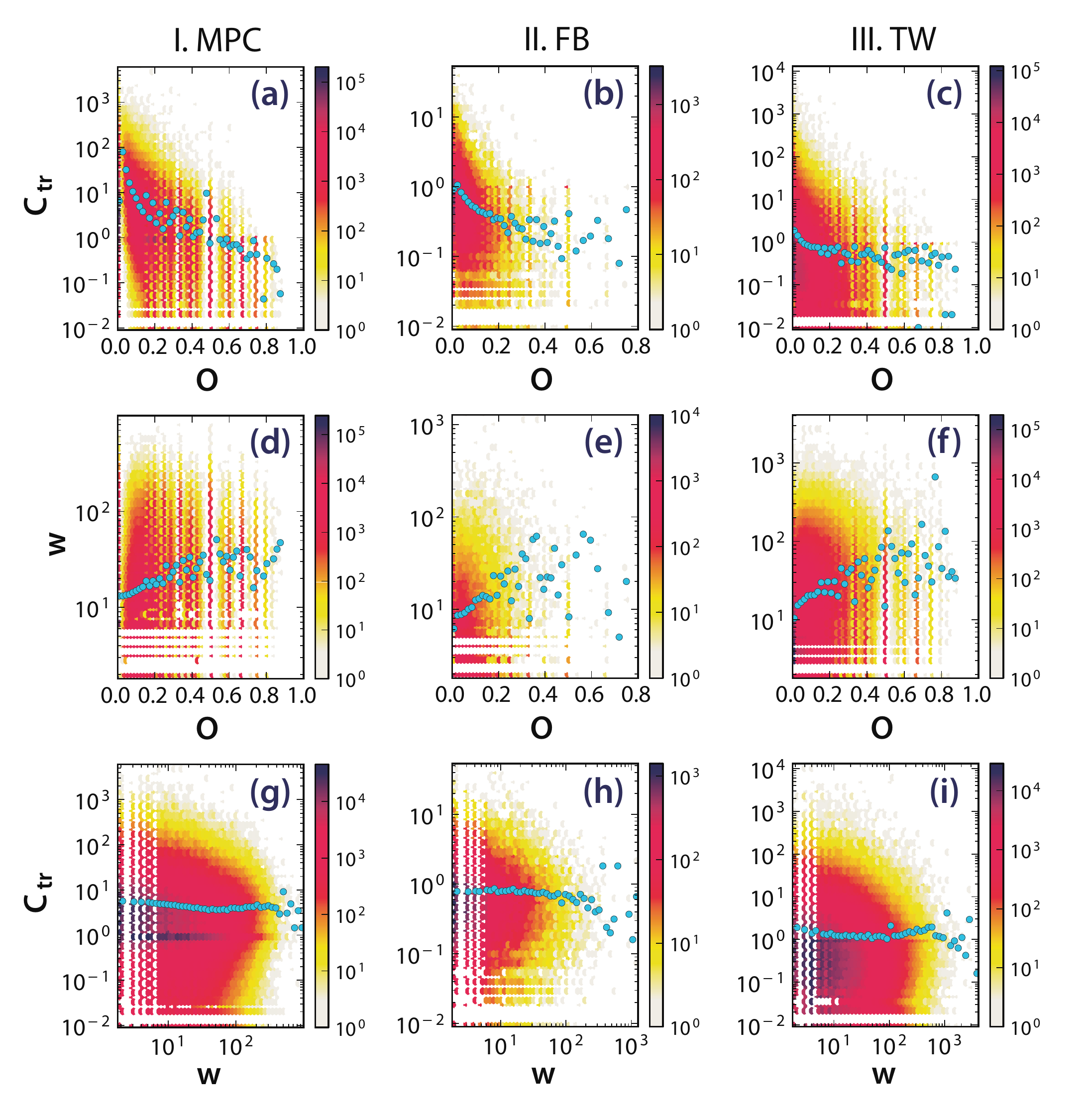}
		\caption{Correlations between transmission centrality $C_{tr}$ and overlap $O$ (a-c), dyadic tie strength $w$ and overlap $O$ (d-f), and between transmission centrality $C_{tr}$ and dyadic tie strength $w$ values (g-i) of links in the MPC (a, d, g), FB (b, e, h), and TW (c, f, i) networks. The colors of the binned heat maps assign the distribution of links with the given pair of tie strength values, while blue symbols show their average over the $y-axis$ as a function of the $x-axis$ values (quantitative measures of correlations between link properties can be found in Table~\ref{tab:correlations}).}% For MPC, the Pearson correlation $r_{O,w}=0.097 (p$-value < $10^{-6}), r_{O, C_{ctr}}=-0.1275 (p$-value < $10^{-6}), r_{w, C_{tr}}=-0.023 (p$-value < $10^{-6})$; for FB, the Pearson correlation $r_{O,w}=0.15 (p$-value < $10^{-6}), r_{O, C_{ctr}}=-0.15 (p$-value < $10^{-6}), r_{w, C_{tr}}=-0.01 (p$-value < $10^{-4})$; for TW, the Pearson correlation $r_{O,w}=0.102 (p$-value < $10^{-6}), r_{O, C_{ctr}}=-0.021 (p$-value < $10^{-6}), r_{w, C_{tr}}=-0.002 (p$-value < $10^{-3})$}
		\label{fig:weaktiecorrelation}
\end{figure}

As shown in Fig.\ref{fig:weaktiecorrelation}d-f and in other studies~\cite{Onnela2007a, Onnela2007b}), dyadic tie strength and link overlap are positively correlated in accordance with Granovetter's theorem~\cite{Granovetter73}. At the same time, transmission centrality and overlap show strong negative correlations (see Fig.\ref{fig:weaktiecorrelation}.a-c) as weak links, with small overlap values, are commonly situated between communities, and thus transmitting information to a large set of nodes. More interestingly, dyadic tie strength and transmission centrality values do not show strong correlations (see in Fig.\ref{fig:weaktiecorrelation}.g-i). Although both are correlated with link overlap, they capture notably different and seemingly independent features of social ties. For the precise Pearson correlation coefficients (and $p$-values) see Table~\ref{tab:correlations}.

While overlap has been shown to identify weak ties efficiently~\cite{Onnela2007a, Onnela2007b}, this measure has a major limitation. It assigns a zero overlap vale for an unrealistically large fraction of links including weak ties but also leaf links, links surrounded by structural holes, or links situated at sparsely connected parts of the network. It is indeed true in the investigated systems where $48.2\%$, $49.8\%$, and $45.2\%$ of social ties appear with $O=0$ (resp. in the MPC, FB, TW networks). Relying merely on the link overlap one cannot differentiate between these links, thus they are treated equivalently. On the other hand, the Granovetterian criteria suggest that weak ties are not only characterized by small overlap, but they also exhibit small dyadic tie strengths, and high transmission centrality. Based on these conditions we design two combined strategies where we differentiate between zero overlap links using their $w$ or $C_{tr}$ values. We first rank ties in an increasing order of overlap, and then sort again links of the same overlap value increasingly by their dyadic tie strength (assigned as $(O,w)$), or by their inverse transmission centrality values (assigned as $(O,C_{tr}^{-1})$). 

\subsubsection{Controlled SIR spreading.}
The precise identification of the weakest weak ties is important, because by suppressing interactions on this limited set of links, we may effectively control globally spreading processes in the network. To model such scenarios we take a network structure and introduce a weight $\omega_{ij}$ for each link (with values defined later). To select the weakest links to control, we consider one of the two candidate sorting strategies, $(O,w)$ or $(O,C_{tr}^{-1})$. After sorting links by one of these metrics, we select the $f$ weakest fraction of them to control by linearly rescaling their weights as $\Omega_{ij}=\delta \omega_{ij}$, with the parameter $0\leq \delta \leq 1$.

\begin{figure}[h!]
\begin{center}
\includegraphics[width=0.75\textwidth]{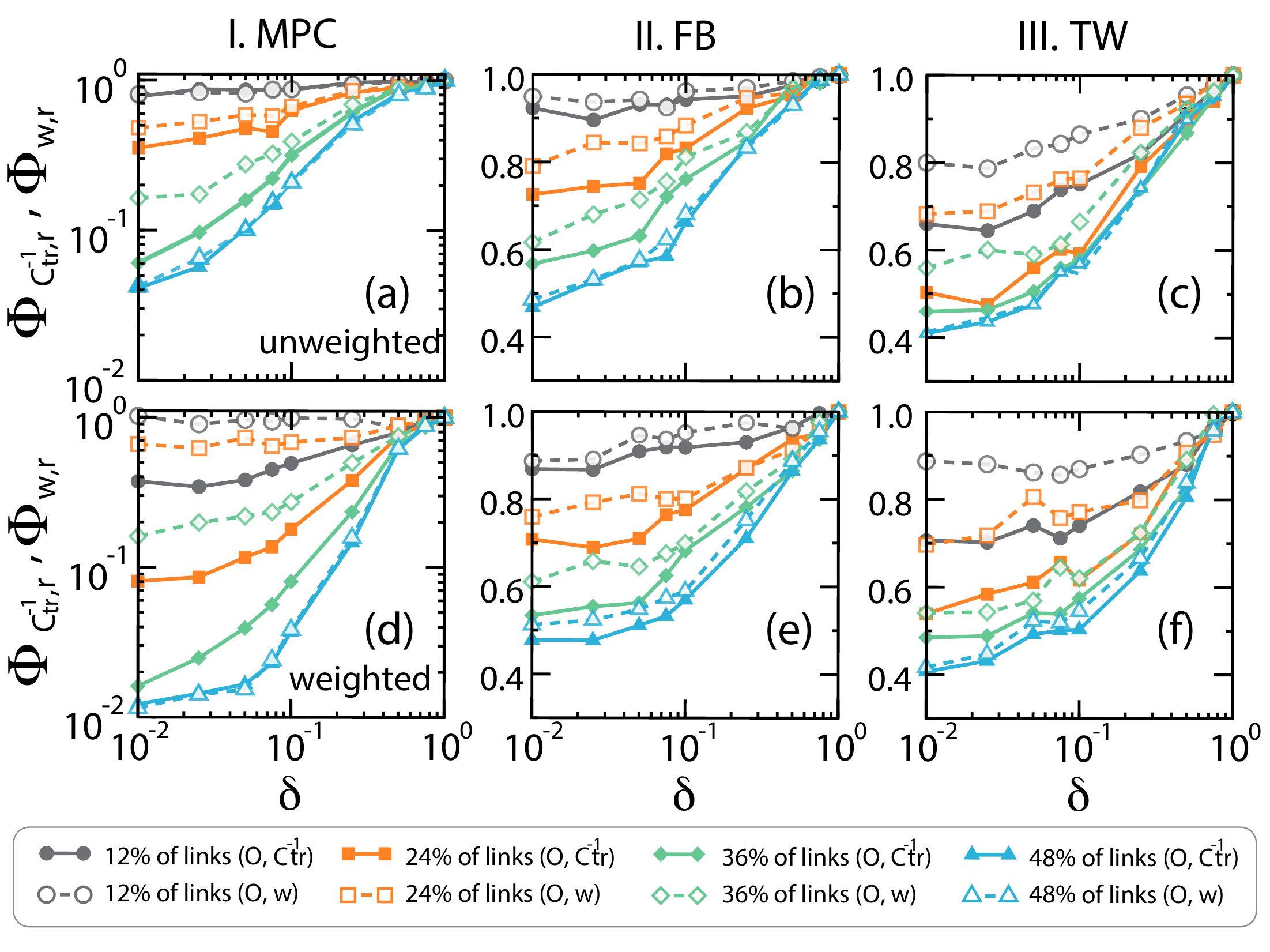}
 \end{center}
 \caption{Controlled SIR spreading experiments. Two ratios $\Phi_{C_{tr}^{-1},r}(\delta)=R_{O,C_{tr}^{-1}}(\delta)/R_{rand}(\delta)$ (solid line) and $\Phi_{w,r}(\delta)=R_{O,w}(\delta)/R_{rand}(\delta)$ (dashed line) of final recovered population sizes are shown as the function of $\delta$ rescaling parameter in case of $(O,C_{tr}^{-1})$, $(O,w)$ and random control strategies when we control the $12\%$ (gray), $24\%$ (orange), $36\%$ (green), $48\%$ (blue) of links. We present results for unweighted  (a) MPC, (b) FB, (c) TW and weighted (d) MPC, (e) FB, (f) TW networks. Results were obtained averaging over SIR spreading processes with parameters $\beta=0.25$ and $\mu=0.1$ initiated from $1000$ random seeds.}
  \label{fig:SIRstudy}
\end{figure}

In this way, we weaken interactions on the selected ties, and such that we can exert further control on dynamical processes, like the Susceptible-Infected-Removed (SIR) model. The SIR process~\cite{Barrat2008} is a well known model of epidemics and rumor spreading \cite{AndersonMay1992,Nekovee2007} and it is defined on a network where nodes can be in exclusive states of susceptible (S), infected (I), or recovered (R) \cite{Barrat2008}. At each iteration connected nodes are updated as $S+I \overset {\beta} {\rightarrow} 2I $, or $I \overset {\mu} {\rightarrow} R$ with $\beta$ and $\mu$ being the infection and recovery rates respectively. In this scenario, we fix $\mu=0.1$ and $\beta=0.25$, and re-scale the transmission probability for each controlled link as $\widetilde{\beta}_{ij}=\Omega_{ij}\beta$ (for a sensitivity analysis regarding this choice see SI Fig. S3). After initiating the process from a randomly selected seed we simulate it until full recovery and monitor $R$, the number of recovered nodes giving the maximum number of nodes ever got infected during the process.

In our first experiment we assign $\omega_{i,j}=1$ for each link assuming that the network is unweighted at the outset. To study the effects of link control, after sorting links by $(O,C_{tr}^{-1})$ or $(O,w)$, we choose the weakest $12\%$, $24\%$, $36\%$, or $48\%$ of links (see Fig.\ref{fig:SIRstudy}.a, b, and c). In addition, as a reference we use a network where the same fraction of randomly selected links are controlled in the same way, i.e., by re-scaling their weights with $\delta$. Finally to quantify the effects of increasing control, we measure the $\Phi_{C_{tr}^{-1},r}(\delta)=R_{O,C_{tr}^{-1}}(\delta)/R_{rand}(\delta)$, and $\Phi_{w,r}(\delta)=R_{O,w}(\delta)/R_{rand}(\delta)$ ratios of recovered nodes in scenarios of targeted and random control strategies for various $\delta$ values. If the targeted strategy performs performs comparable to the random one, these ratios are equal to one; otherwise the stronger control a targeted strategy enforces, the smaller the corresponding ratio becomes.

When we set $\delta=1$ the ratios of endemic population size are trivially one as no control is applied (see Fig.\ref{fig:SIRstudy}.a, b, and c). However by decreasing $\delta$, thus by increasing control, large differences appear between the targeted and random cases. Effects are stronger when a larger fraction of weakest links are re-scaled with smaller and smaller $\delta$ factor. The differences between the $(O,C_{tr}^{-1})$ (solid lines) and $(O,w)$ (dashed lines) strategies are maximal when we control an intermediate $24\%$ or $36\%$ of links, while they perform similarly when the controlled fraction is small ($12\%$) or large ($48\%$). It is also evident that the $(O,C_{tr}^{-1})$ strategy outperforms the $(O,w)$ and provides remarkably better control in reducing the final infected population, specially for smaller $\delta$ values.

\begin{figure}[ht!]
\begin{center}
\includegraphics[width=0.75\textwidth]{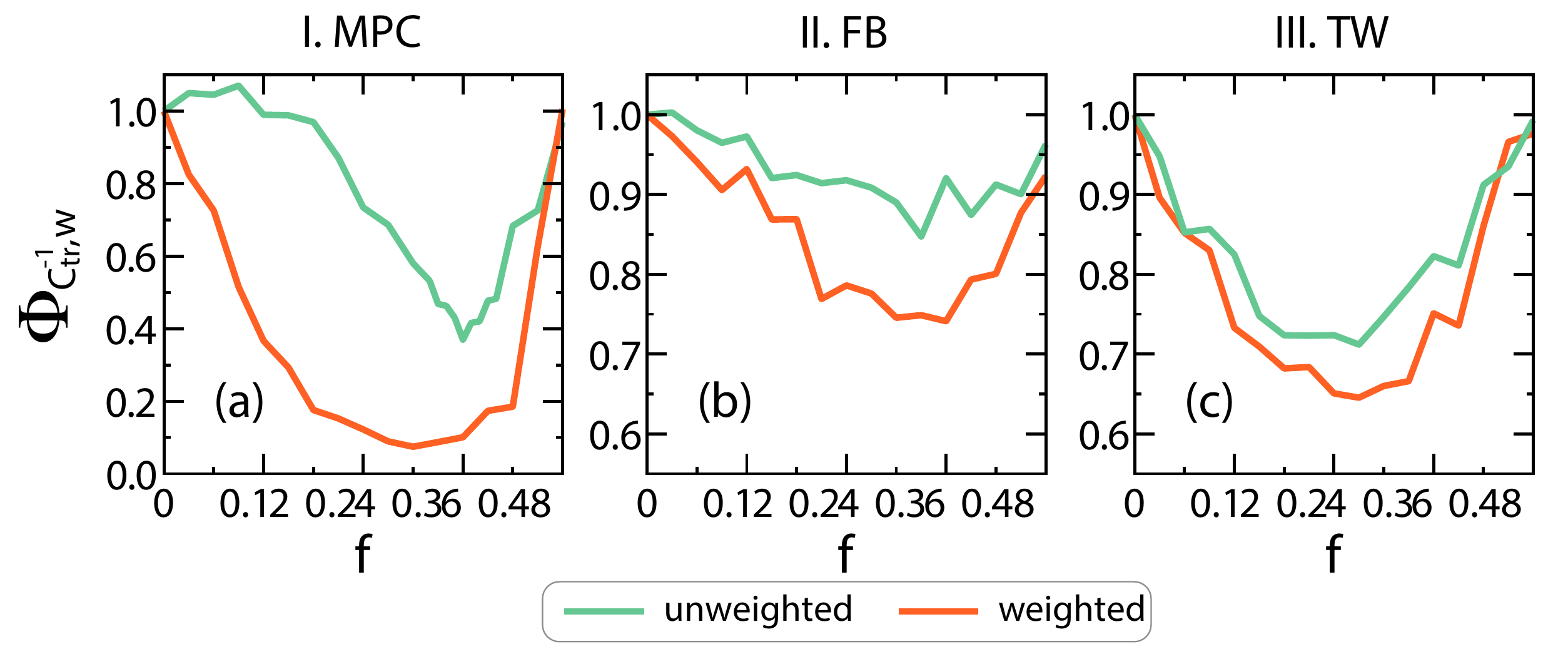}
 \end{center}
 \caption{The $\Phi_{C_{tr}^{-1},w}(f)=R_{O,C_{tr}^{-1}}(\delta)/R_{O,w}(\delta)$ ratio of endemic recovered population sizes of $(O,C_{tr}^{-1})$ and $(O,w)$ strategies as a function of controlled $f$ fractions of links. Calculations are repeated in unweighted (green) and weighted (red) networks of (a) MPC, (b) FB, and (c) TW. Results are obtained averaging over SIR spreading processes with parameters $\beta=0.25$, $\mu=0.1$ and $\delta=0.01$ initiated from $1000$ random seeds.}
  \label{fig:SIRrate}
\end{figure}

To bring our experiments closer to reality we repeat our measurements on weighted networks where we define link weights as $\omega_{ij}=w_{ij}/\langle w \rangle$, i.e. the number of interactions between nodes $i$ and $j$ normalized by the $\langle w\rangle$ average number of interactions per link calculated over the whole network. In the case where $\omega_{ij} > \langle w \rangle$ we set the corresponding weight $\omega_{ij}=1.0$. This choice is necessary as weights are heterogeneously distributed in this case, and thus severely slow down the simulated spreading to reach full prevalence. On the other hand, since controlled links with small overlap values tend to have small weights, negligible effect of this approximation is expected. The different control strategies qualitatively provide the same results on the weighted FB and TW networks (Fig.\ref{fig:SIRstudy}.e, f); however, their effects are considerably stronger on the MPC structure (Fig.\ref{fig:SIRstudy}.d). There, the $(O,C_{tr}^{-1})$ strategy appears to be the more efficient even after controlling only the $12\%$ of the ties. Moreover, this strategy can lead to $90\%$ reduction of the infected population in the case of re-scaling $36\%$ of links with $\delta=0.01$. Note that the observed differences between different strategies cannot be the result of the limited communication on zero overlap links only, as we observed qualitatively the same effects in weighted and unweighted networks.

To directly highlight the differences between the targeted strategies we further investigate the strongest controlled case. We set $\delta=0.01$ and repeat our experiments by controlling various $f$ fractions of links to measure the $\Phi_{C_{tr}^{-1},w}(f)=R_{O,C_{tr}^{-1}}(\delta)/R_{O,w}(\delta)$ fraction of endemic recovered population sizes, i.e., the ratio of the two performance functions. Results in Fig.\ref{fig:SIRrate}.a, b, and c evidently show that the $(O,C_{tr}^{-1})$ strategy almost always outperforms the $(O,w)$ strategy, especially when we consider weights. In addition, the minimum points of the $\Phi_{C_{tr}^{-1},w}(f)$ curves in Fig.\ref{fig:SIRrate} assign the best pay-off between the controlled $f$ fraction of links and the effectiveness of contamination control using the $(O,C_{tr}^{-1})$ strategy. This minimum point indicates that $\sim 30\%$ of the weakest ties are enough to control and mostly efficiently hinder the spreading processes on the investigated social networks.

\section{Discussion}

In this study we introduced a new link centrality measure, called \textit{transmission centrality}, which sensitively quantifies the importance of links in global diffusion processes. We defined an algorithm to compute transmission centrality, demonstrated on three large-scale networks its general properties, and discussed possible ways of how this measure can be generalized for directed, weighted or temporal networks or even as a node centrality measure. Finally in a case study, we showed that the combined information of overlap and transmission centrality serves as the best way to identify weak links to gain maximum control of spreading processes. Although here we demonstrated the effectiveness of transmission centrality in identifying weak ties in social networks specifically, the same metric can be applied in any other type of networks to identify links with specific structural role and importance in controlling the emergence of various collective phenomena.

We discussed that the main limitation of this new centrality measure is rooted in its computational complexity, which scales as the best known algorithm for betweenness centrality. However, we proposed a way around this limitation by defining a heuristic method to approximate transmission centrality values in very large networks at a considerably cheaper cost.

Several extensions of this method are possible by considering other probing processes other than SI, or arbitrary weight definitions, directed links, temporal interactions, or node transmission centrality. Furthermore, several straightforward applications can be foreseen. Examples are in viral marketing, rumor contamination, or intervention designs; their identification can be the subject of future studies. Our aim here is to ground a new metric of link centrality and to contribute to the design of effective methods to identify ties, which play an indisputably important role in the structure and dynamics of social networks.

\section*{Acknowledgements}
  We are grateful for D. Mocanu for her help in data preparation and N. Samay for her help in visualization. We acknowledge the support from the SoSweet (ANR-15-CE38-0011-01) ANR project. QZ would like to acknowledge Dr.~Duygu Balcan for mentorship and invaluable contributions in the beginning of this project.

\bibliography{weaktie}
\newpage


\section*{Supplementary material}

# Supplementary Information:
## *Link transmission centrality in large-scale social networks*

Qian Zhang, Márton Karsai, Alessandro Vespignani

## S1  Local bias of link transmission centrality

In the main text, we have discussed that link transmission centrality is a locally biased measure as it assigns higher values to links, which are closer to the seed node. To understand this bias on the "close-to-seed" links, we directly analyze the actual branching trees with root as the actual seed node. We compute the transmission centrality $C_{tr}$ of links as the function of their distance $d$ from the actual root. Here the distance of a link is defined as $d = \min(\ell_b, \ell_e)$, i. e., the minimum of the shortest paths $\ell_b$ or $\ell_e$ of the beginning and ending nodes (respectively) of the actual link.

One can assume two different characters for links in the vicinity of the seed node. First, if the seed has low degree, the corresponding $C_{tr}$ values of immediate links will be larger compared to the case where the seed has many neighbors. Second, we expect that this bias decreases by distance $d$ measured from the seed. These two effects can be identified from the scaling of the $C_{tr}(d)$ functions in Fig.S1.a-c. Here for each network we select randomly seeds from different degree groups (100 seeds for each group) and measure the $C_{tr}$ average transmission centrality (Fig.S1.a-c) in distance smaller than or equal to $d$ relative to the actual seed.

These results can help us estimate the induced bias and select an appropriate threshold $d_{max}$ to eliminate its effect. This choice has to consider two competing factors: the choice of a $d_{max}$, which is large enough that the local bias of the seed becomes negligible; and to choose a distance small enough not to remove too many links from the tree. The $E$ number of links in distance $d$ from the seed is exponentially increasing as shown in (Fig.S1.d-f). To remove the actual bias we set $C_{tr} = 0$ for those links, which are within a determined distance $d_{max}$ from the actual seed in the network. Based on Fig.S1 we choose $d_{max} = 8$, $d_{max} = 3$, and $d_{max} = 7$, for the MPC, FB and TW datasets respectively. These choices fulfill both conditions as they are large enough to reduce the bias considerably, while at the same time exclude only the 4.3%, 1.1% and 3.7% of links for MPC, FB and TW datasets respectively. Naturally, removing links may decrease the overall heterogeneity of $C_{tr}$. However, as demonstrated in Fig.S1.g-i, even after excluding the biased links, the distribution $P(C_{tr})$ remains fat tailed.

## S2  Relation to betweenness centrality

Transmission centrality as a measure can be easily associated to the concept of link betweenness centrality commonly defined as $C_b(i,j) = \sum_{s \neq i \neq j \neq t} g_{s,t}(i,j)/g_{s,t}$, where $g_{s,t}$ assigns the number of shortest paths between nodes $s$ and $t$ while $g_{s,t}(i,j)$ is the number of them, which goes through the link $(i,j)$. Although

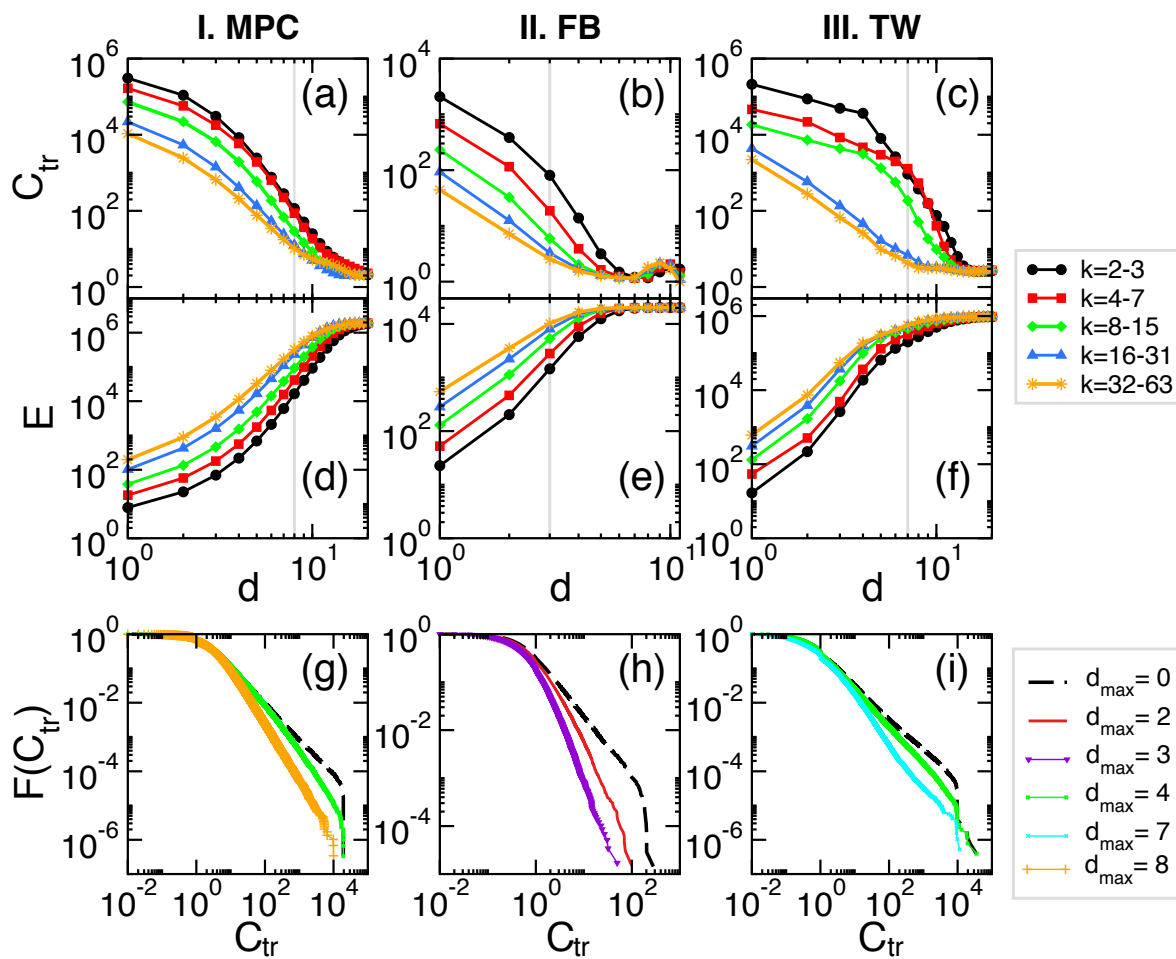

Figure S1: Local bias of link transmission centrality on the MPC, the FB and the TW networks. (a,b,c) Average importance $C_{tr}$ of links in distance d from the actual seed for the three network respectively. (d,e,f) Average number of links in distance d from the actual seed. Random seeds were selected from different degree groups (100 seeds for each group). (g,h,i) Distribution of transmission centrality values after unbiasing with links in distance d.

the definition of $C_b$ and $C_{tr}$ are not equivalent they are strongly related. Their differences are rooted in the deterministic definition of $C_b$, which considers all shortest paths between pairs of nodes. On the other hand, $C_{tr}$ is defined by an SI process, which is stochastic even in case of $\beta = 1$, as it considers in a random order the possible links of an infected node along which to transmit the infection to susceptible neighbours. In this way it may never explore *all possible shortest paths* but would give credit to the most plausible ones over several attempts of realizations. Despite these fundamental differences they both capture similar quantities fractional to the number of shortest paths running through a given link. Due to these underlying similarities

they appear to be closely correlated in the case of all investigated empirical structures as seen in Fig.S2.

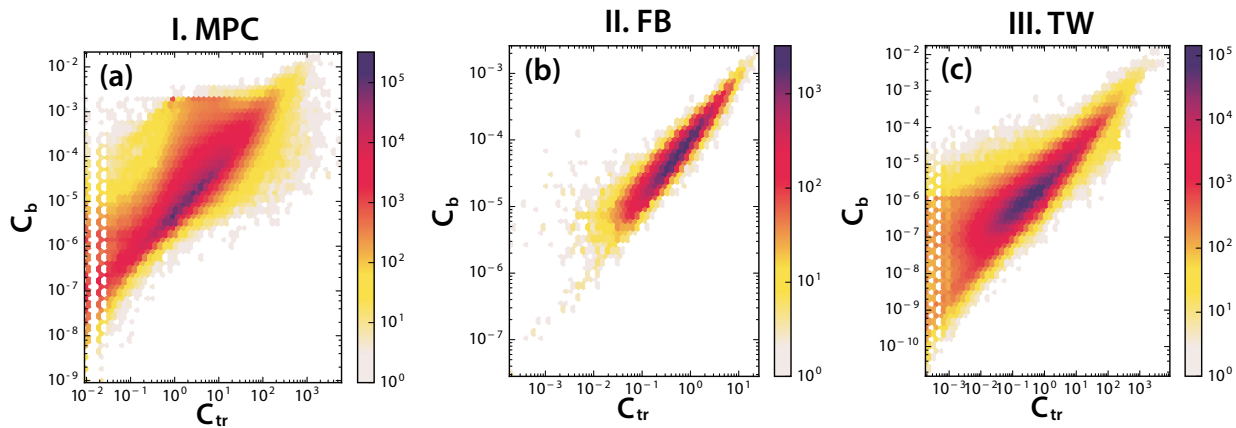

Figure S2: (a,b,c) Correlations between sampled and unbiased link transmission centrality $C_{tr}$ and link betweenness centrality $C_b$ values in case of the MPC (Pearson $r = 0.83$, p-value $< 10^{-6}$), FB (Pearson $r = 0.97$, p-value $< 10^{-6}$), and TW (Pearson $r = 0.91$, p-value $< 10^{-6}$) networks respectively. Computations of $C_{tr}$ were initiated from 2000 (5000 and 5000) seeds in each case (respectively). Orange symbols assign average $C_b$ values for each $C_{tr}$. For the MPC network $C_b$ values were calculated between 5000 randomly selected nodes and any other nodes in the network.

These strong correlations demonstrate the relationship between transmission centrality and betweenness centrality, with the advantage that the approximate calculation of the former scales considerably better with system size. As discussed earlier, the exact computation of transmission centrality scales with $\mathcal{O}(|V||E|)$ complexity, which is equal to the best known algorithm [?] to measure $C_b$. On the other hand, one can reduce radically the computational cost by considering a relatively small number of seeds to compute average $C_{tr}$ and to obtain surprisingly good approximations for link betweenness values. To demonstrate this scaling, for all three empirical networks we measured the correlations between $C_b$ and $C_{tr}$ while successively increasing the number of seeds for the latter one. Results summarized in Table S1 show that the average transmission centrality, computed from the 0.1% of the MPC network nodes (10% in case of FB and 0.2% for TW), approximates well the actual betweenness centrality values, with correlations $R_{MPC} \simeq 0.83$ (resp. $R_{FB} \simeq 0.96$ and $R_{TW} \simeq 0.87$). This demonstrates yet again the close relationship between these metrics and the success of the provided approximation method of transmission centrality.

## S3 Parameter dependence of controlled SIR spreading

In the main text we argued that the combined metrics of overlap-transmission centrality $(O, C_{tr}^{-1})$ is the most efficient metric to hinder epidemics outbreaks modeled by an SIR process. However, all presented simulation results used a single parameter set with a constant basic reproduction number of $R_0 = \beta/\mu = 2.5$. Here, to demonstrate that our simulation results were mostly independent of the choice of $R_0$, we fixed $\mu = 0.1$ and repeated our experiments for different values of $\beta$. We selected an f fraction of links by the

Table S1: The r Pearson correlation coefficients measured between the betweenness centrality and transmission centrality values of the same links with increasing number of seeds in the three empirical network. Each corresponding p-value is smaller than 0.001.

| No. Seed | 10 | 50 | 100 | 500 | 1000 | 2000 |
|---|---|---|---|---|---|---|
| $r_{MPC}$ | 0.16 | 0.27 | 0.48 | 0.67 | 0.79 | 0.83 |
| $r_{FB}$ | 0.36 | 0.58 | 0.74 | 0.92 | 0.94 | 0.96 |
| $r_{TW}$ | 0.40 | 0.55 | 0.65 | 0.76 | 0.84 | 0.87 |

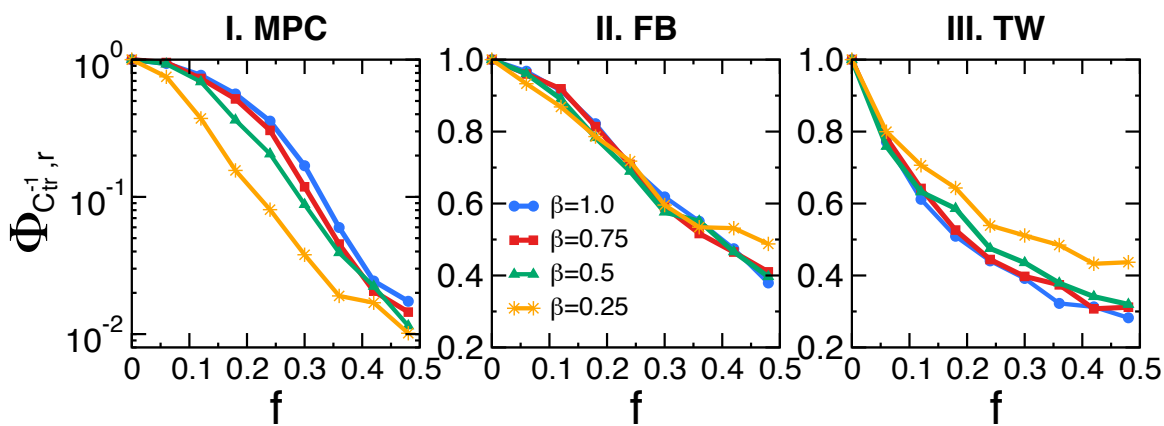

Figure S3: Parameter dependence of SIR spreading. We control weak ties to impede SIR spreading on (a) the MPC, (b) the FP, and (c) the TW networks with fixed scaling factor $\delta = 0.01$. The recovery rate $\mu$ of the SIR process is fixed to 0.1. We show the $\Phi_{C_{tr}^{-1},r}(f) = R_{O,C_{tr}^{-1}}(f)/R_{rand}(f)$ ratio between the endemic recovered population sizes after scaling the weights of links selected by the $(O, C_{tr}^{-1})$ strategy and randomly. We simulate the process for different values of the infection rate $\beta$ (1.0 blue dots, 0.75 red squares, 0.5 green triangles, and 0.25 orange stars) as the function of the f fraction of controlled links.

$(O, C_{tr}^{-1})$ strategy or randomly, and re-scaled their weight by $\delta = 0.01$. We measured the average final size of the recovered population through 100 realizations for the targeted and random strategies. Depicting the ratio $\Phi_{C_{tr}^{-1},r}(f) = R_{O,C_{tr}^{-1}}(f)/R_{random}(f)$ of the corresponding measures, in Fig.S3, we show that the effects of targeted control increases as we control a higher fraction of links (just as we have seen in the main text); however this behaviour depends only weakly on the choice of $\beta$. This suggests that the observed behaviour is qualitatively the same for a wide range of $R_0$ of the SIR model, thus not a consequence of specific parametrization of the spreading process.


\end{document}